\documentclass[aps,showpacs,amsmath,amssymb,twocolumn,prl,floatfix,superscriptaddress]{revtex4}
\usepackage[dvips]{graphicx}

\begin{document}
\bibliographystyle{apsrev}

%
\title{The nonrelativistic limit of Dirac-Fock codes: the role of Brillouin configurations}

\author{ P. Indelicato}
\email{paul@spectro.jussieu.fr}
\affiliation{
Laboratoire Kastler Brossel, \'Ecole Normale Sup\'erieure et Universit\'e P. et M. Curie, Case 74, 4 place Jussieu, F-75252, Cedex 05, France}
\author{ E. Lindroth}
\affiliation{Atomic Physics, Fysikum, Stockholm University, S-106
91 Stockholm, Sweden}
\author{J.P. Desclaux}
\affiliation{15 Chemin du Billery, 38360 Sassenage}
\date{\today}
\begin{abstract}
We solve a long standing problem with relativistic calculations done
with the widely used Multi-Configuration Dirac-Fock Method (MCDF).
We show, using Relativistic Many-Body Perturbation Theory (RMBPT),
how even for relatively high-$Z$, relaxation or correlation causes
the non-relativistic limit of states of different total angular
momentum but identical orbital angular momentum to have different
energies. We  show that only large scale calculations that include
all single excitations, even those obeying the Brillouin's theorem
have the correct limit. We reproduce very accurately recent
high-precision measurements in F-like Ar, and turn then into precise
test of QED. We obtain the correct non-relativistic limit not only
for fine structure but also for level energies and show that RMBPT
calculations are not immune to this problem.
\end{abstract}
 \maketitle

Relativistic atomic structure codes, mostly MCDF packages,
 are now of widespread use in many sectors of physics, and the need for reliable,
relativistic calculations is stronger than ever (see, e.g.,
\cite{beiersdorfer:2003} for examples in Astrophysics). However, the
difficulties of doing reliable calculations are numerous, and still
largely underestimated. For example a puzzle that was noted already
twenty-two years ago~\cite{huang:82:rlimit}  has never been solved,
although  it may lead even in very simple calculations to wrong
energy values. In Ref.~ \cite{huang:82:rlimit} it was shown that
relativistic self-consistent field procedures do not produce, in a
number of cases, the correct non-relativistic limit of  zero for the
fine structure splitting (FSS) when the speed of light is tuned to
infinity. Ref.~\cite{huang:82:rlimit} suggested  as a remedy
explicit calculation  of this {\em non-relativistic offset} (N.R.)
and subsequent  subtraction of it from the relativistic result,
although no justification for the procedure was provided.
 Moreover this paper said nothing on how to correct individual energy levels.
 Here we will penetrate the origin of the non-relativistic shift using the tools of
perturbation theory and advanced MCDF calculations. We use these
tools to show the role of relaxation in the N.R. offset, and prove
that the inclusion of  specific mono-excitations in the MCDF basis
removes it. We also provide justification to the subtraction
procedure and show that not  only the FSS need to be corrected, but
also the level energy, e.g., when transitions between different
shells are studied. It is also worth noting that this problem
appears in the Optimized Level (OL) scheme when each level energy
and wavefunction is optimized separately. This scheme is used only
when the highest accuracy for correlation is required. Often the
average  (AL) level scheme is used, in which the \emph{same}
$J$-average wavefunction is used to calculate the energy of all FS
component. In the AL scheme the N.R. offset does not appear, but the
accuracy is much lower.

We will concentrate on the ground state configuration of a F-like
ion which was used as a model system already in
Ref.~\cite{huang:82:rlimit} as accurate measurements have been
performed very recently \cite{draganic:03}. With high experimental
accuracy, even for $Z=18$,  it is important to be aware of this
problem which seriously affects the comparison with experiment on
the present day level. We will further present accurate calculations
of the fine structure splitting in F-like argon both with
Relativistic Many-Body Perturbation Theory (RMBPT) and with the
Multi-Configuration Dirac-Fock (MCDF) method. It is shown that by
comparison with accurate experimental results~\cite{draganic:03} it
is possible to test the calculations on self-energy and other
radiative corrections in a true many-electron surrounding.

With RMBPT  the fine-structure splitting in a F-like system is
calculated as the binding energy difference  between the $2p_{1/2}$
and $2p_{3/2}$ electron in the corresponding Ne-like system. The
lowest order approximation of this binding energy
 is the negative of the orbital energy of the removed
electron in the Hartree-Fock approximation. The remaining electrons
are at this stage considered as {\em frozen} in their orbitals in
spite of the removal of one electron. The most important correction
to this first approximation is the {\em relaxation} of the electrons
due to the presence of the hole. The term {\em relaxation} usually
denotes the correction found by a single configuration restricted
Hartree-Fock (or {\em Dirac-Fock} in the relativistic case)
calculation in the presence of the hole. The non-relativistic shift
has its origin already at this level and we will now concentrate on
this shift and postpone the discussion of higher order corrections.

\begin{figure}
\includegraphics[width=\columnwidth,clip=true,trim= 3cm 25cm 10cm 0cm,angle=0]{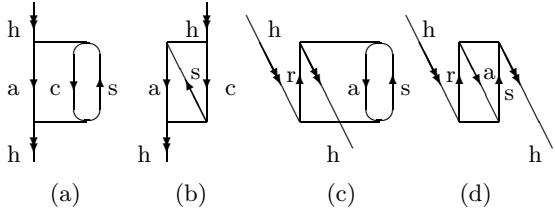}
 \caption{Illustration of the second order energy
contributions to a one-hole system. Diagrams(a-b) show fluctuations
to two holes and one excited orbital and diagrams (c-d) double
excitations (correlation). Downgoing single arrows denote core
orbitals, downgoing double arrows denote the hole and upgoing arrows
denote excited orbitals.} \label{fig:piccorr}
\end{figure}

To analyze the relaxation for a one-hole state with perturbation
theory it is natural to start from the  closed shell system and
systematically correct for the removal of one electron.
Fig.~\ref{fig:piccorr} shows the contributions entering in second
order. Fig.~\ref{fig:piccorr}(a-b) show fluctuations to two holes
and one excited orbital and Fig.~\ref{fig:piccorr}(c-d) true double
excitations. The {\em relaxation}, i.e., the effects included by a
single configuration restricted Hartree-Fock calculation is in
perturbation theory part of Fig.~\ref{fig:piccorr}(a-b); the ones
where the hole is not fluctuating and the excitation from an orbital
preserves its angular symmetry.
The lowest order relaxation correction to an orbital $b$ can
consequently be written
\begin{equation}
\label{eq:relax} \rho^{relax}_b  \left(\ell_s = \ell_b, j_s = j_b
\right) = - \sum_s^{exc} \frac{\mid s \rangle \langle \left\{ hs
\right\} \mid V_{12} \mid \left\{ hb \right\}
\rangle}{\varepsilon_b-\varepsilon_s}
\end{equation}
where $h$ denotes the removed electron, the curly brackets
antisymmetrization, $V_{12}$ the two-electron interaction, and the
minus sign is due to the removal of $h$.
 The energy corrections are then calculated as
\begin{equation}
\label{eq:relaxE} \sum_b^{core}
 \langle  \left\{ bh \right\} \mid V_{12}\mid  \left\{ \rho^{relax}_b  h
 \right\}\rangle.
\end{equation}
In this way all types of diagrams in Fig.~\ref{fig:piccorr}~(a-b)
with either orbital $a=h$ (and b=c) or $c=h$ (and b=a) and $\ell_s =
\ell_b$, $ j_s = j_b$ are included, i.e., the single excitations
that preserve the angular structure.  It can be noted that these
single excitation contributions form a class of diagrams that can be
summed until convergence in an iterative scheme, see, e.g.,
Refs.~\cite{mooney:92:xe,indelicato:92:ka}. Here we will not pursue
this line, however, since our purpose is to analyze the relaxation
in the non-relativistic limit and show why a state with a hole in
$n\ell_{j=\ell-1/2}$ and one with a hole $n\ell_{j=\ell+1/2}$ do not
reach the same energy in this limit. For this it is sufficient to
study relaxation in second order.

As an example, take diagram Fig.~\ref{fig:piccorr}(a) with orbital
$a=h$ and $\ell_s=\ell_b$, a typical relaxation contribution. The
orbitals used to evaluate the diagram are solved using the
Hartree-Fock potential from the closed shell core and the radial
part of the $2p_{1/2}$ and the $2p_{3/2}$ orbital will be
identical when we let $c \rightarrow \infty$. The problem comes
instead from the spin-angular part. Since
\begin{equation}
\mid \ell m_{\ell} \,  s m_{s} \rangle = \sum_{jm_j} \mid \left(
\ell s \right) j m_{j} \rangle \langle  \left( \ell s \right) j
m_{j} \mid \ell m_{\ell} \, s m_{s} \rangle
\end{equation}
decoupling of spin and orbital angular momentum cannot be done
without summing over all total angular momenta, $j$. An
unambiguous way to see how this influences our example of
Fig.~\ref{fig:piccorr}(a) with orbital $a=h$ and $\ell_s=\ell_b$ is
to compare the angular contribution non-relativistically and
relativistically. The electron-electron interaction is expressed
as
\begin{equation}
\frac{1}{r_{12}}=\sum_k \frac{r_<^k}{r_>^{k+1}} \textbf{C}^k\left(
1 \right) \cdot \textbf{C}^k\left( 2 \right),
\end{equation}
where $k$ denotes the rank of the spherical tensor operator
$\textbf{C}$, which works on the orbital part of wavefunctions. Non
relativistically the angular part can be evaluated as
\begin{equation}
\label{eq:nonrel} \sum_b^{core} \sum_k 2 \frac{1}{2k+1} \,
\frac{1}{2\ell_h+1} \langle \ell_h \mid  \mid \textbf{C}^k \mid
\mid \ell_h \rangle^2 \langle \ell_b \mid  \mid \textbf{C}^k \mid
\mid \ell_b \rangle^2.
\end{equation}
This is in fact identical to the following expression in the
coupled space where two extra sums appear over intermediate total
angular momenta
\begin{eqnarray}
\label{eq:rel} \sum_b^{core} \sum_k \sum_{j_{h'}}^{\ell_h \pm
\frac{1}{2}} \, \, \sum_{j_{b'}}^{\ell_b \pm \frac{1}{2}}
\frac{1}{2k+1} \, \frac{1}{2 j_h+1} \, \,
\nonumber \\
\langle j_{h}  \mid \mid \textbf{C}^k \mid \mid j_{h'} \rangle^2
%
\langle j_{b}  \mid  \mid \textbf{C}^k \mid \mid j_{b'} \rangle^2,
\end{eqnarray}
That these two expressions give the same result can be understood by standard angular momentum algebra techniques.

In a restricted Dirac-Fock calculation there will be no sums over
intermediate angular momenta. Instead only $j_h=j_{h'}$ is
allowed, i.e., the hole is not allowed to fluctuate to the other
fine structure component, and $j_{b'}=j_b$ is required , i.e., the
corrections to orbital $b$ do not change its angular structure.
 The spin-angular part used is thus
\begin{eqnarray}
\label{eq:rel2} \sum_b^{core} \sum_k \frac{1}{2k+1} \,
\frac{1}{2j_h+1} \, \,  \langle j_{h}  \mid \mid \textbf{C}^k \mid
\mid j_{h} \rangle^2 \langle j_{b}  \mid  \mid \textbf{C}^k \mid
\mid j_{b} \rangle^2,
\end{eqnarray}
which will clearly not produce the same result as
Eq.~\eqref{eq:nonrel}, and which further cannot give identical
results for, e.g., $j_h=1/2$ and $j_h=3/2$, which is easily seen
from the $k=2$ contribution which is zero for $j_h=1/2$, but not for
$j_h=3/2$. The difference can also be readily demonstrated
numerically for a system as F-like neon where  the second order
contribution, Eq.~\ref{eq:relaxE},  to the relaxation gives an
unphysical fine structure offset of $0.024$ eV in the $c
\longrightarrow \infty$ limit. Following the recipe from
Ref.~\cite{huang:82:rlimit} and correcting the result calculated
with the true value of $c$ with this offset, we obtain a relaxation
contribution to the  fine structure splitting of $-0.058$~eV. After
iteration of the relaxation
contributions~\cite{mooney:92:xe,indelicato:92:ka} the corrected
value reaches $ \sim -0.050$~eV,  in line with the MCDF Coulomb
relaxation contribution of $ \sim -0.049$~eV,
 listed in the third section of Table \ref{tab:summary}. This value  has been corrected using the same
 procedure. The small difference is probably due to  small differences
 in the classification of relaxation and correlation
 contributions.
The lesson here is that since the
  summation over all possible couplings of spin and orbital
  angular momenta
 of the intermediate states are necessary to reproduce the uncoupled
 situation a correct non-relativistic limit
cannot be  achieved with any single configuration self consistent
field calculation. In other words, still for the system under consideration, one has to include
 more than one configuration relativistically  to reproduce the single
 configuration non-relativistic result in a relativistic framework.
 In RMBPT the full contribution from Fig.~\ref{fig:piccorr}(a-b)
 produces no N.R. offset, but any attempt to speed up the convergence
 of the perturbation expansion by singling out the important
 subclass that only involve energy denominators as in
 Eq.~\ref{eq:relax} will do so.
 The RMBPT results shown in Table \ref{tab:summary} are obtained without any such procedure
 and has a correct non-relativistic limit by
construction.

\begin{table}
\squeezetable
\caption{\label{tab:summary} Summary of the contributions. All calculations use the 2002 values for fundamental constants \cite{mohr:codata:00,mohr:04} (eV). Experimental values are from wavelength provided in Ref.~\cite{draganic:03} converted to vacuum values using \cite{stone:04}}
\begin{ruledtabular}
\begin{tabular}{lrrr}
&\multicolumn{1}{c}{$2p_{1/2}$}& \multicolumn{1}{c}{$2p_{3/2}$} & \multicolumn{1}{c}{$\Delta$} \\
\hline
\multicolumn{4}{c}{ Contributions} \\
Ne-like DF orb. ener. & 426.50002 & 424.13211 & 2.36791 \\
$\Delta$ DF-Breit & -0.22659 & -0.13576 & -0.09083 \\
h.o. retardation & -0.00011 & 0.00079 & -0.00090 \\
QED corr. & 0.01353 & 0.00755 & 0.00598 \\
\hline
\multicolumn{4}{c}{Contributions specific to RMBPT} \\
2nd order
core-core, Coul & -4.48509 & -4.42587 & -0.05921 \\
core-core, Breit & -0.01187 & -0.00814 & -0.00373 \\
correlation, Coul & 2.56726 & 2.55763 & 0.00962 \\
correlation, Breit & 0.02391 & 0.02018 & 0.00373 \\
h.o. contr. (Coul.+Breit) & 0.16559 & 0.15885 & 0.00674 \\
$\Delta$ DF Breit orbitals & 0.00198 & 0.00043 & 0.00156 \\
\hline
Total (RMBPT) & 424.54863 & 422.30777 & 2.24086 \\
Experiment & & & 2.24010 \\
\hline
\multicolumn{4}{c}{Contributions specific to MCDF (N.R. offset subtracted)} \\
Relaxation (Coul) & -3.10800 & -3.05931 & -0.04869 \\
Relaxation (Breit) & -0.00406 & -0.00314 & -0.00092 \\
Correlation (Coul  $\to 5g$) & 1.42466 & 1.39604 & 0.02862 \\
Correlation (Breit  $\to 5g$) & -0.01359 & 0.00741 & -0.02100 \\
\hline
Total (MCDF) & 424.58585 & 422.34569 & 2.24016 \\
Experiment & & & 2.24010 \\
\end{tabular}
\end{ruledtabular}
\end{table}


 With several configurations included it should in principle be
 possible to reach the correct non-relativistic limit, in
 practice one can, however, generally not achieve this in a {\em truncated}
 calculation. In practice the number of configurations has to be
 truncated for all but the smallest systems. It is  common to truncate after double
 excitations from the dominating configuration, but just as double excitations are needed to be added to
the single excitations to obtain the correct non-relativistic limit,
triple excitations will be needed to be added to corresponding
double excitations and so on. Since higher multiple excitations are
less important the remaining offset will however decrease steadily.

\begin{table}[tb]
\squeezetable
 \caption{ \label{tab:mcdf} Contributions to the MCDF energy affected by the N.R. offset (eV).
``$\Delta E$ doub. Exc. $\to\, n=i$'' : correlation
 energy for the configuration space which include all double
  excitations up to principal quantum number $n=i$. Rel. Val.: Relativistic Value. N.R. Off.: Offset obtained at the non-relativistic limit.}
\begin{ruledtabular}
\begin{tabular}{lrrr}
 & \multicolumn{1}{c}{Rel. Val.} & \multicolumn{1}{c}{N.R. Off.} & \multicolumn{1}{c}{Diff.} \\
 \hline
Dirac-Fock Coulomb & 2.31626 & -0.00148 & 2.31774 \\
\multicolumn{4}{c}{Brillouin single excitations excluded} \\
$\Delta E$  Exc. $\to\, n=3$ & -0.01855 & -0.02086 & 0.00231\\
$\Delta E$  Exc. $\to\, n=4$ & -0.01421 & -0.01926 & 0.00505\\
$\Delta E$  Exc. $\to\, n=5$ & -0.01641 & -0.02247 & 0.00606\\
Total  & 2.21621 & -0.02395 & 2.24016 \\
Diff. With Exp. & -0.02389 & & 0.00006 \\
\hline
\multicolumn{4}{c}{All single and double excitations included} \\
$\Delta E$  Exc. $\to\, n=3$ & -0.00371 & -0.00582 & 0.00211 \\
$\Delta E$  Exc. $\to\, n=4$ & 0.00445 & -0.00037 & 0.00482 \\
$\Delta E$  Exc. $\to\, n=5$ & 0.00661 & 0.00075 & 0.00586 \\
Total (S.E.S. Welton) & 2.23923 & -0.00073 & 2.23996 \\
Diff. With Exp. & -0.00087 & & -0.00014 \\
\end{tabular}
\end{ruledtabular}
\end{table}

\begin{figure}
\centering
\includegraphics[height=6.cm,clip=true,trim=0.05cm 0.1cm 0.1cm 0.cm,angle=-90]{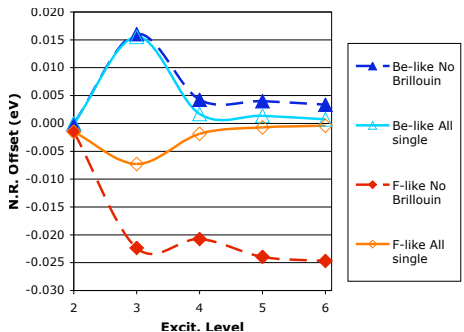}
 \caption{Comparison of the non-relativistic offset  for Be-like and F-like argon, evaluated
 including all single excitations, or only those not obeying Brillouin's theorem.} \label{fig:nroff}
\end{figure}

We now proceed to demonstrate the vanishing of the non-relativistis
offset in an essentially complete MCDF calculation. In the present
calculation we have added to the original configuration all single
and double excitations up to a given maximum $n$ and $\ell$. Note
that one has to be careful in considering the meaning of single and
double excitations. For example the $1s^2 2s^{2} 2p^{4} 3p$  is a
single excitation in the $LS$ coupling sense. Yet in $jj$ coupling
it gives rise to 5 configurations in the $J=1/2$ case, two of which
are double excitations in the $jj$ sense ($2p_{1/2}2p_{3/2}^{4}\to
2p_{1/2}^{2} 2p_{3/2}^{2} 3p_{1/2}$ and $2p_{1/2}^{2} 2p_{3/2}^{2}
3p_{3/2}$).We went from $3d$ to $5g$ for the case with a normal
speed of light, and up to $6h$ for the non-relativistic limit. This
represents respectively  299, 1569, 4339 and 9127 fully relaxed $jj$
configurations for the $J={1/2}$ case, and 456, 2541, 7356 and 15915
for the $J=3/2$. The calculations are repeated with different lists
of configurations. In one group of calculations, we  include all
single and double excitation in the $jj$ sense,  except for the
"Brillouin  single excitations", i.e., those that should contribute
only in third order, as stated by Brillouin's theorem
\cite{bauche:72,godefroid:1987:brillouin,froese:00}. These
excitations are often excluded since they complicate the numerical
convergence. Again we use here Brillouin's theorem in the $jj$
sense, i.e., we exclude all configurations transformed from the
initial one by replacing an orbital with quantum numbers $n$,
$\kappa$ by one with $n'$, $\kappa$, where $\kappa$ is the Dirac
angular number. In a second group we include all single and double
excitations. In both groups, we do calculations  once with only the
Coulomb interaction between electrons used in the evaluation of
wavefunctions and energies, and once with  the full Breit
interaction in the evaluation of wavefunctions and mixing
coefficients. This allows to include high-orders of the Breit
interaction in the calculation. In each group the Coulomb only
calculation is done also a second time with a large value for the
speed of light. The evolution of the N.R. shift as a function of the
maximum excitation used in the MCDF process is plotted in
Fig.~\ref{fig:nroff},
 for both F-like and Be-like ions, to show the generality of what is observed: the N.R. offset
tends to a non-zero constant value when Brillouin configurations are excluded, and to zero when
all single excitations are included.

 The different contributions to the MCDF result,
and the variation of the correlation energy and non-relativistic
offset with and without Brillouin configurations are presented in
Table~\ref{tab:mcdf}. When comparing both results, it is clear that
excluding Brillouin  single excitations and then subtracting the
N.R. offset leads to the same result as including the Brillouin
configurations. The agreement with experiment and with RMBPT results
is excellent in both cases, even though the quality of the
convergence when including all single excitations is not as good as
when Brillouin ones are excluded. Moreover, the inclusion of all
single excitations enables also to correct the energy of a level as
shown in Table \ref{tab:ground-corr}, which was not possible with
the subtraction method. Finally we note that the evaluation of the
radiative corrections, the self-energy screening (SES) with the help
of the Welton approximation \cite{Indelicato:1987:helike} leads to a
very good agreement with experiment.

\begin{table}[tb]
\squeezetable
 \caption{ \label{tab:ground-corr} Change in the ground state ($J=3/2$) correlation energy due to Brillouin single excitation}
\begin{ruledtabular}
\begin{tabular}{lrrr}
Conf. & No-brillouin & All single & Diff \\
\hline
corr $\to 3d$ & -5.1792 & -5.1989 & -0.0196 \\
corr $\to 4f$ & -7.7349 & -7.7603 & -0.0255 \\
corr $\to 5g$ & -8.6551 & -8.6871 & -0.0320 \\
\end{tabular}
\end{ruledtabular}
\end{table}

In conclusion we have proven, by comparing RMBPT and MCDF results,
that the N.R. offset is due to relaxation and should go away when
doing a complete calculation. We then showed that in the MCDF case,
the offset is going to zero  if a large enough configuration space
is used, but only if \emph{all single} configurations are included.
In practice  excluding Brillouin single excitations and then
subtracting the N.R. offset leads to the same value, but numerical
convergence of the self-consistent field process is much easier in
the latter case. Finally, failing to account for the N.R. offset
leads to poor results, even at a moderately large $Z$, a fact that
may not have received enough attention in many MCDF calculations.
The present work also shows that similar problems can happen in
RMBPT calculations if subclasses of important effects are singled
out and by themselves are treated to higher order. The improved
convergence will then come at the expense of an N.R. offset. This
fact had not been recognized before.

Laboratoire Kastler Brossel is Unit{\'e} Mixte de Recherche du
CNRS n$^{\circ}$ 8552. Financial support for this research was
received from the Swedish Science Research Councils (VR).

\bibliography{nonrellimit}

\begin{thebibliography}{12}
\expandafter\ifx\csname natexlab\endcsname\relax\def\natexlab#1{#1}\fi
\expandafter\ifx\csname bibnamefont\endcsname\relax
  \def\bibnamefont#1{#1}\fi
\expandafter\ifx\csname bibfnamefont\endcsname\relax
  \def\bibfnamefont#1{#1}\fi
\expandafter\ifx\csname citenamefont\endcsname\relax
  \def\citenamefont#1{#1}\fi
\expandafter\ifx\csname url\endcsname\relax
  \def\url#1{\texttt{#1}}\fi
\expandafter\ifx\csname urlprefix\endcsname\relax\def\urlprefix{URL }\fi
\providecommand{\bibinfo}[2]{#2}
\providecommand{\eprint}[2][]{\url{#2}}

\bibitem[{\citenamefont{Beiersdorfer}(2003)}]{beiersdorfer:2003}
\bibinfo{author}{\bibfnamefont{P.}~\bibnamefont{Beiersdorfer}},
  \bibinfo{journal}{Annu. Rev. Astron. Astrophys.}
  \textbf{\bibinfo{volume}{41}}, \bibinfo{pages}{343} (\bibinfo{year}{2003}).

\bibitem[{\citenamefont{Huang et~al.}(1982)\citenamefont{Huang, Kim, Cheng, and
  Desclaux}}]{huang:82:rlimit}
\bibinfo{author}{\bibfnamefont{K.~N.} \bibnamefont{Huang}},
  \bibinfo{author}{\bibfnamefont{Y.~K.} \bibnamefont{Kim}},
  \bibinfo{author}{\bibfnamefont{K.~T.} \bibnamefont{Cheng}}, \bibnamefont{and}
  \bibinfo{author}{\bibfnamefont{J.~P.} \bibnamefont{Desclaux}},
  \bibinfo{journal}{Phys. Rev. Lett.} \textbf{\bibinfo{volume}{48}},
  \bibinfo{pages}{1245} (\bibinfo{year}{1982}).

\bibitem[{\citenamefont{Draganic et~al.}(2003)\citenamefont{Draganic, {Crespo
  L\'{o}pez-Urrutia}, DuBois, Fritzsche, Shabaev, {Soria Orts}, Tupitsyn, Zou,
  , and Ullrich}}]{draganic:03}
\bibinfo{author}{\bibfnamefont{I.}~\bibnamefont{Draganic}},
  \bibinfo{author}{\bibfnamefont{J.~R.} \bibnamefont{{Crespo
  L\'{o}pez-Urrutia}}},
  \bibinfo{author}{\bibfnamefont{R.}~\bibnamefont{DuBois}},
  \bibinfo{author}{\bibfnamefont{S.}~\bibnamefont{Fritzsche}},
  \bibinfo{author}{\bibfnamefont{V.~M.} \bibnamefont{Shabaev}},
  \bibinfo{author}{\bibfnamefont{R.}~\bibnamefont{{Soria Orts}}},
  \bibinfo{author}{\bibfnamefont{I.~I.} \bibnamefont{Tupitsyn}},
  \bibinfo{author}{\bibfnamefont{Y.}~\bibnamefont{Zou}}, , \bibnamefont{and}
  \bibinfo{author}{\bibfnamefont{J.}~\bibnamefont{Ullrich}},
  \bibinfo{journal}{Phys. Rev. Lett.} \textbf{\bibinfo{volume}{91}},
  \bibinfo{pages}{183001} (\bibinfo{year}{2003}).

\bibitem[{\citenamefont{Mooney et~al.}(1992)\citenamefont{Mooney, Lindroth,
  Indelicato, Kessler, and Deslattes}}]{mooney:92:xe}
\bibinfo{author}{\bibfnamefont{T.}~\bibnamefont{Mooney}},
  \bibinfo{author}{\bibfnamefont{E.}~\bibnamefont{Lindroth}},
  \bibinfo{author}{\bibfnamefont{P.}~\bibnamefont{Indelicato}},
  \bibinfo{author}{\bibfnamefont{E.~G.} \bibnamefont{Kessler}},
  \bibnamefont{and} \bibinfo{author}{\bibfnamefont{R.~D.}
  \bibnamefont{Deslattes}}, \bibinfo{journal}{Phys. Rev. A}
  \textbf{\bibinfo{volume}{45}}, \bibinfo{pages}{1531} (\bibinfo{year}{1992}).

\bibitem[{\citenamefont{Indelicato and Lindroth}(1992)}]{indelicato:92:ka}
\bibinfo{author}{\bibfnamefont{P.}~\bibnamefont{Indelicato}} \bibnamefont{and}
  \bibinfo{author}{\bibfnamefont{E.}~\bibnamefont{Lindroth}},
  \bibinfo{journal}{Phys. Rev. A} \textbf{\bibinfo{volume}{46}},
  \bibinfo{pages}{2426} (\bibinfo{year}{1992}).

\bibitem[{\citenamefont{Mohr and Taylor}(2004)}]{mohr:04}
\bibinfo{author}{\bibfnamefont{P.~J.} \bibnamefont{Mohr}} \bibnamefont{and}
  \bibinfo{author}{\bibfnamefont{B.~N.} \bibnamefont{Taylor}}
  (\bibinfo{year}{2004}),
  \bibinfo{note}{http://physics.nist.gov/cuu/Constants/index.html}.

\bibitem[{\citenamefont{Mohr and Taylor}(2000)}]{mohr:codata:00}
\bibinfo{author}{\bibfnamefont{P.~J.} \bibnamefont{Mohr}} \bibnamefont{and}
  \bibinfo{author}{\bibfnamefont{B.~N.} \bibnamefont{Taylor}},
  \bibinfo{journal}{Rev. Mod. Phys.} \textbf{\bibinfo{volume}{72}},
  \bibinfo{pages}{351} (\bibinfo{year}{2000}).

\bibitem[{\citenamefont{Stone and Zimmerman}(2004)}]{stone:04}
\bibinfo{author}{\bibfnamefont{J.~A.} \bibnamefont{Stone}} \bibnamefont{and}
  \bibinfo{author}{\bibfnamefont{J.~H.} \bibnamefont{Zimmerman}}
  (\bibinfo{year}{2004}), \bibinfo{note}{http://emtoolbox.nist.gov/}.

\bibitem[{\citenamefont{Bauche and Klapisch}(1972)}]{bauche:72}
\bibinfo{author}{\bibfnamefont{J.}~\bibnamefont{Bauche}} \bibnamefont{and}
  \bibinfo{author}{\bibfnamefont{M.}~\bibnamefont{Klapisch}},
  \bibinfo{journal}{J. Phys. B: At. Mol. Phys.} \textbf{\bibinfo{volume}{5}},
  \bibinfo{pages}{29} (\bibinfo{year}{1972}).

\bibitem[{\citenamefont{Froese~Fischer
  et~al.}(2000)\citenamefont{Froese~Fischer, Brage, and J\"onsson}}]{froese:00}
\bibinfo{author}{\bibfnamefont{C.}~\bibnamefont{Froese~Fischer}},
  \bibinfo{author}{\bibfnamefont{T.}~\bibnamefont{Brage}}, \bibnamefont{and}
  \bibinfo{author}{\bibfnamefont{P.}~\bibnamefont{J\"onsson}},
  \emph{\bibinfo{title}{Computational Atomic Structure}}
  (\bibinfo{publisher}{Institute of Physics Publishing},
  \bibinfo{address}{Bristol}, \bibinfo{year}{2000}).

\bibitem[{\citenamefont{Godefroid et~al.}(1987)\citenamefont{Godefroid, Lievin,
  and Metz}}]{godefroid:1987:brillouin}
\bibinfo{author}{\bibfnamefont{M.}~\bibnamefont{Godefroid}},
  \bibinfo{author}{\bibfnamefont{J.}~\bibnamefont{Lievin}}, \bibnamefont{and}
  \bibinfo{author}{\bibfnamefont{J.~Y.} \bibnamefont{Metz}},
  \bibinfo{journal}{J. Phys. B: At. Mol. Phys.} \textbf{\bibinfo{volume}{20}},
  \bibinfo{pages}{3283} (\bibinfo{year}{1987}).

\bibitem[{\citenamefont{Indelicato et~al.}(1987)\citenamefont{Indelicato,
  Gorceix, and Desclaux}}]{Indelicato:1987:helike}
\bibinfo{author}{\bibfnamefont{P.}~\bibnamefont{Indelicato}},
  \bibinfo{author}{\bibfnamefont{O.}~\bibnamefont{Gorceix}}, \bibnamefont{and}
  \bibinfo{author}{\bibfnamefont{J.}~\bibnamefont{Desclaux}},
  \bibinfo{journal}{J. Phys. B: At. Mol. Opt. Phys.}
  \textbf{\bibinfo{volume}{20}}, \bibinfo{pages}{651} (\bibinfo{year}{1987}).

\end{thebibliography}

\end{document}